\documentclass[aps, twocolumn,noshowpacs,preprintnumbers,amsmath,amssymb]{revtex4}

\usepackage{verbatim}
\usepackage{wasysym}

\usepackage{graphicx}
\usepackage{dcolumn}
\usepackage{bm}
\usepackage{wrapfig}

\begin{document}

\title{Influence of the deposition conditions on the field emission properties of patterned nitrogenated carbon nanotube films}

\author{Jean-Marc Bonard}
\author{Ralph Kurt}
\author{Christian Klinke}
\affiliation{Departement de Physique, Ecole Polytechnique Federale de Lausanne, CH-1015 Lausanne EPFL, Switzerland}

\begin{abstract} 

The electron field emission of patterned films of nitrogenated carbon nanotubes is shown to be decisively influenced by the deposition conditions. The growth was carried out by decomposing methane in a nitrogen/ammonia atmosphere using hot filament chemical vapor deposition (CVD). The diameter of the produced tubes depended critically on the distance between substrate and filament, and the field emission properties  applied fields needed for electron emission, field amplification factor) could be directly correlated to the film morphology. This demonstrates the possibility of tuning the field emission properties of such film emitters. For arrays of nanotubes thinner than 50 nm, the onset of field emission was observed at $\sim$ 4 V/$\mu$m, and a current density of 10 mA/cm$^{2}$ was obtained for an applied field of 7.2 V/$\mu$m.
\end{abstract}

\maketitle

\section*{Introduction}

Field emission has emerged as one of the most promising applications for carbon-based films, as attested by an increasing effort from researchers all over the world. In particular, carbon nanotube emitters \cite{1,2} are purported to be ideal candidates for the next generation of field emission flat panel displays \cite{3}. Carbon
nanotubes are capable of emitting high currents (up to 1 A/cm$^{2}$) at low fields ($\sim$ 5 V/$\mu$m) \cite{4}. Furthermore, the controlled deposition of nanotubes on a substrate has recently become possible through the combined use of chemical vapor deposition  CVD) methods \cite{5,6} and catalyst patterning techniques (e.g., laser patterning \cite{7}, standard lithography \cite{8,9}, soft lithography \cite{10}, or self-assembly \cite{11}).

The challenge now is to take part of these exceptional properties to realize devices at the industrial scale. One critical issue at present is a better control of the growth, in order to optimize the emission and to ensure reproducible characteristics. In fact, recent studies have stressed the determinant influence of the film morphology on the emission properties (e.g. \cite{12,13}). It is therefore very important to gain a better understanding of the influence of various deposition parameters. We demonstrate in this contribution how the morphology, and hence the field emission properties, of patterned films of nitrogenated carbon (C:N) nanotubes grown by hot filament-CVD (HF-CVD) on Si substrates can be controlled with the deposition conditions.

\section*{Growth of patterned C:N nanotube films}

\begin{figure}[ht]
  \centering
  \includegraphics[width=0.45\textwidth]{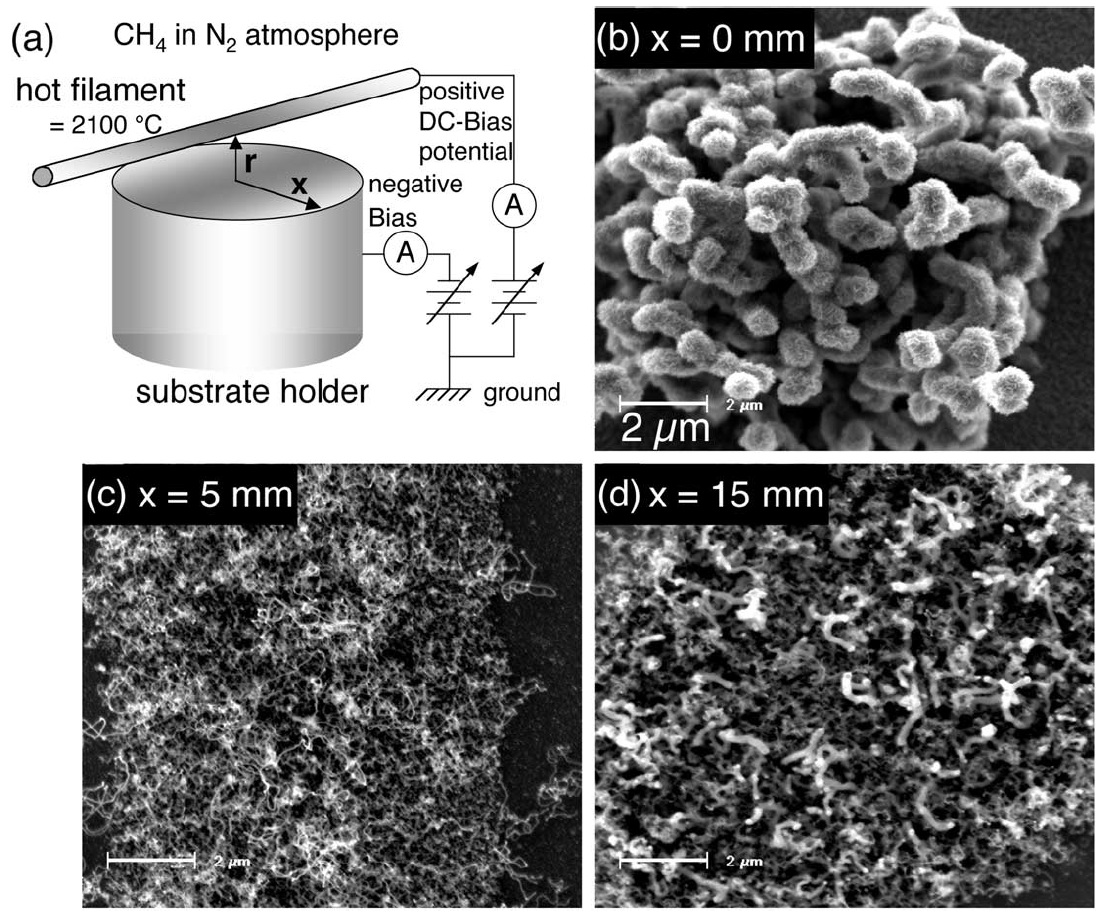}
  \caption{\textit{(a) Schematics of the experimental setup, and  (b) - (d) SEM micrographs showing the dependence of the film morphology as a function of the lateral distance $x$ (and hence of the substrate temperature $T_{sub}$): (b) $x = 0$ mm ($T_{sub} = 820^{\circ}$C); (c) $x = 5$ mm ($T_{sub} = 740^{\circ}$C); (d) $x = 15$ mm ($T_{sub} = 720^{\circ}$C).}}
\end{figure}

Recently, we demonstrated the possibility of growing C:N nanotubes \cite{14} by decomposition of CH$_{4}$ in a N$_{2}$ and NH$_{3}$ atmosphere by bias-enhanced HF-CVD \cite{15,16}. The experimental setup is shown in Fig. 1a and consists in one resistively heated WC filament (diameter 500 $\mu$m, filament temperature $T_{fil} = 2100 ^{\circ}$C) placed at a distance $r = 5$ mm over a Mo sample holder. In this configuration, the overall distance between filament and substrate varies with the lateral distance $x$. This provokes a gradual variation of the deposition conditions, as the temperature, composition and flux of the gas mixture as well as the temperature of the substrate $T_{sub}$ depend on the distance to the filament. For example, $T_{sub}$ was measured with a W-5\%Re/W-26\%Re thermocouple, and was found to decrease from 780$^{\circ}$C directly underneath the filament down to 690$^{\circ}$C at $x = 25$ mm for a working pressure of 300 Pa (also shown in
Fig. 4a). The total gas flow was kept constant at 500 sccm and a typical flow ratio of N$_{2}$:CH$_{4}$:NH$_{3}$ of 100:1:1 was applied. The deposition time was 30 min.

Although C:N nanotubes can be deposited on a pure Si substrate \cite{16}, the presence of a catalyst allows one to control the growth and to define patterns of nanotubes on the substrate \cite{17}. Best results were obtained using microcontact printing ($\mu$CP) \cite{10} to transfer an ethanolic ink containing 60 mM Fe(NO$_{3}$)$_{3}$ $\cdot$ 9H$_{2}$O and 20 mM Ni(NO$_{3}$)$_{2}$ $\cdot$ 6H$_{2}$O to the Si substrate \cite{17}. The resulting structures are depicted in the scanning electron microscopy (SEM) pictures of Fig. 1b-d. All tubes are bent or coiled, and show lengths of 10-50 $\mu$m as estimated from the micrographs. The diameter of the nanotubes was found to vary between 30 nm and 1 $\mu$m as a function of $x$ (see also Fig. 4b). The thickest tubes were found directly underneath the
filament, as shown in Fig. 1b. The diameter decreased then with increasing $x$ down to $\sim$ 50 nm for $x = 5$ mm (Fig. 1c). For larger distances, we found a further decrease in diameter for some tubes while other tubes showed an increase in diameter, resulting in a wide dispersion (see Fig. 1d for $x = 15$ mm).

Fig. 2 shows the resulting structures for $x = 5$ mm (corresponding to Fig. 1c) as observed by transmission electron microscopy (TEM) \cite{16}. The nanotubes are hollow and show a typical diameter of 50 nm. All studied tubes were decorated similarly to the tubes shown in Fig. 2a. The needle-like structures protruding from the surface of the tubes are in fact thin bent or rolled-up graphene sheets. The disordered graphitic-like character of the tubes was confirmed by Raman spectroscopy \cite{17}. We could furthermore show that the degree of coiling and decoration depends on the amount of nitrogen incorporated in the tubes, and a maximum N/C atomic ratio of 0:043 $\pm$ 0:014 was obtained according to electron energy loss spectroscopy \cite{16}. We also observed, as in Fig. 2b, that a catalyst particle of complex
shape is present at the tip of the nanotubes. The decoration seen in Fig. 2a is also less pronounced near the tip of the tube, as a feathered structure is found only a few hundred nanometer away from the tip. The end of the tubes are always closed, as high magnification microscopy reveals a crystalline structure, corresponding roughly to the  (002) basal planes of graphite, surrounding the catalyst particle.

\begin{figure}[ht]
  \centering
  \includegraphics[width=0.45\textwidth]{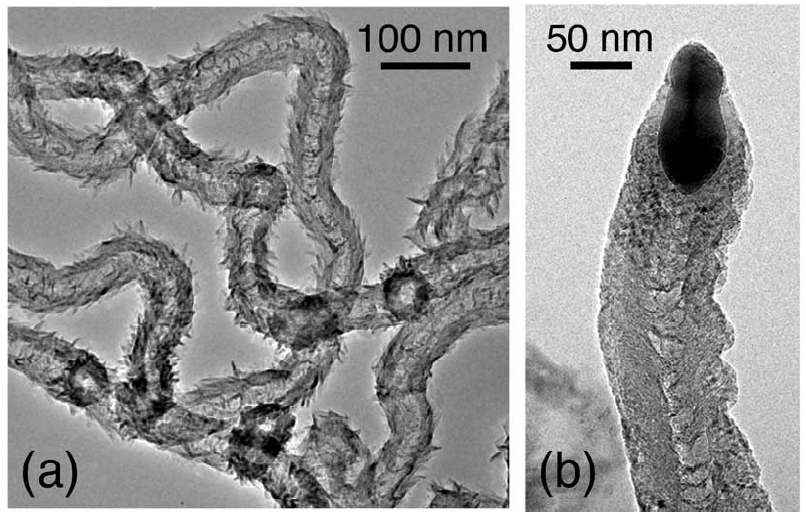}
  \caption{\textit{TEM micrographs of the resulting C:N nanotubes for $x = 5$ mm: (a) overview of the deposit; (b) detail of a nanotube tip.}}
\end{figure}

\section*{Field emission}

A sample such as the one depicted in Figs. 1 and 2 represent a unique opportunity to study in detail the influence of the film morphology on the field emission properties. We characterized the field emission at a base pressure of 10$^{-7}$ mbar by collecting the emitted electrons with a polished stainless steel spherical counterelectrode of 1 cm diameter, which corresponds to an emission area of $\sim$ 0.007 cm$^{2}$. The samples were mounted on a linear manipulator, and the inter-electrode distance d$_{0}$ was adjusted to 125 $\mu$m prior to emission. It was decreased to 50 and finally 10 $\mu$m when no field emitted current was detected below 1100 V applied voltage. We present in the following results obtained on one sample for the sake of clarity, but the same behavior has been observed on several other samples.

\begin{figure}[ht]
  \centering
  \includegraphics[width=0.45\textwidth]{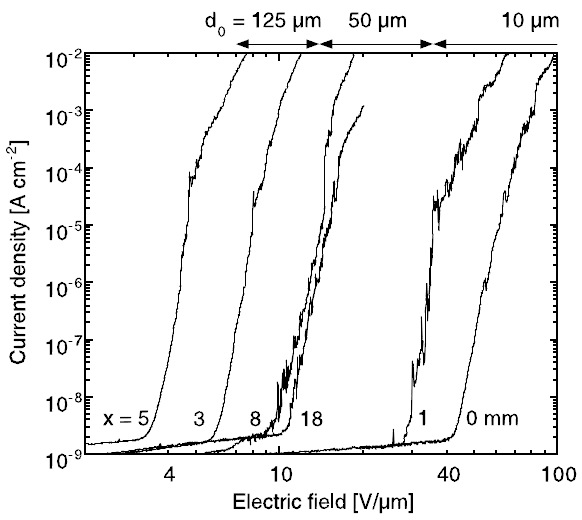}
  \caption{\textit{Field emission $I - V$ characteristics taken at different lateral distances $x$. The applied field is given as the macroscopic field $V/d_{0}$ at the cathode surface, and the interelectrode distance $d_{0}$ is also indicated.}}
\end{figure}

The field emission was measured at 28 locations on the film between $x = 0 - 20$ mm, and Fig. 3 shows clearly a strong variation in the field emission characteristics. Directly underneath the filament ($x \leq 1$ mm), the interelectrode distance had to be decreased to 10 $\mu$m to observe field emission below 1100 V. As $x$ increased, the fields needed for emission decreased rapidly, and reached a minimum for $x = 5$ mm. The emission fields increased again, and levelled off for distances larger than $x = 8$ mm. We found that a current density of 10 mA/cm$^{2}$ could be reached repeatedly at every measured location, and that the emission was stable over a few hours (no longer durations were considered here).

To analyze the results in more detail, we extracted several parameters from the $I - V$ characteristics of Fig. 3 and displayed them in Fig. 4 as a function of the lateral distance $x$. The field amplification factor $\beta$ (Fig. 4c) was obtained by fitting the $I - V$ curves at low emitted currents with the Fowler-Nordheim formula using a workfunction of 5 eV, which is a reasonable assumption for carbon-based field emitters \cite{12,19}. Fig. 4d shows the onset field $E_{i}$, the turn-on field $E_{to}$ and the threshold field $E_{thr}$, which are the electric fields $V/d_{0}$ needed to extract current densities of 10 nA/cm$^{2}$, 10 $\mu$A/cm$^{2}$, and 10 mA/cm$^{2}$, respectively. To correlate these parameters with the film morphology, the temperature on the substrate $T_{sub}$ as well as the tube diameter as estimated by SEM are given in Fig. 4a,b, respectively.

\begin{figure}[ht]
  \centering
  \includegraphics[width=0.45\textwidth]{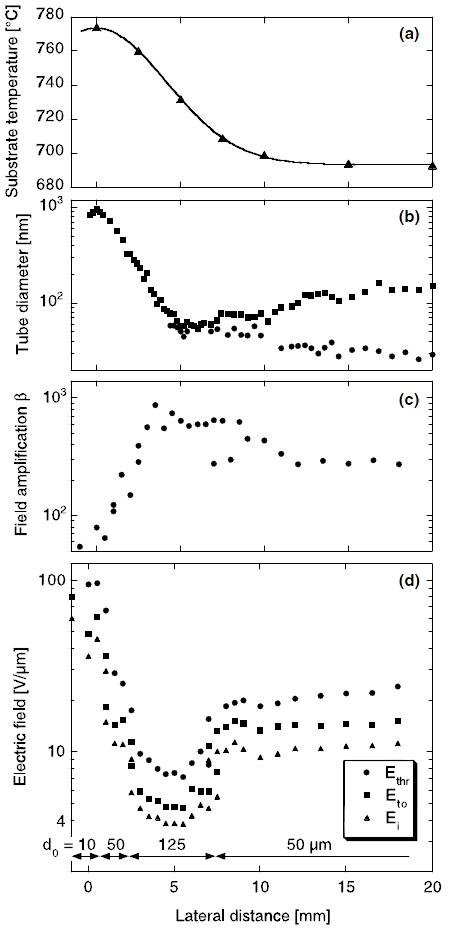}
  \caption{\textit{Structural and field emission parameters as a function of the lateral distance $x$: (a) substrate temperature $T_{sub}$; (b) mean tube diameter as estimated by SEM; (c) field amplification factor $\beta$; (d) onset ($E_{i}$), turn-on ($E_{to}$) and threshold field ($E_{thr}$). The interelectrode distance $d_{0}$ is also indicated. The spread in the fields and $\beta$ from one measurement to the next are typically $\pm$5\% and $\pm$15\%, respectively.}}
\end{figure}

Fig. 4 reveals that the field emission properties of the C:N nanotube film vary over a large range and clearly depend on the lateral distance $x$, and hence on the deposition conditions. The very high emission fields measured directly underneath the filament ($E_{i} = 36$ V/$\mu$m, $E_{thr} = 95$ V/$\mu$m) coincide with a large tube diameter and a correspondingly low field amplification. We observe then a spectacular correlation between these three parameters: as the mean diameter decreases sharply, the emission fields decrease drastically with a corresponding increase in field amplification. The lowest fields were achieved at $x = 5$ mm ($E_{i} = 3.7$
V/$\mu$m, $E_{thr} = 7.2$ V/$\mu$m, corresponding to temperatures between 730$^{\circ}$C and 750$^{\circ}$C and tube diameters of $\sim$ 50 nm.

The morphology of the film becomes more complex for larger $x$, as two nanotube populations appear, one showing a further decrease in diameter down to 30 nm, and the other characterized by a slow increase followed by a saturation around $x = 15$ mm to diameters of $\sim$ 150 nm (Fig. 4b). Interestingly, the emission fields increase and the field amplification decreases beyond $x = 5$ mm: the tendency does not follow the decrease in diameter of the smaller tube population (lower branch in the upper panel of Fig. 4). Far from the filament, typical fields of $E_{i} \approx 10$ V/$\mu$m and $E_{thr} \approx 20$ V/$\mu$m are obtained.

On emitter assemblies like the ones shown in Fig. 1, only the nanotubes with the highest field amplification will emit, which in turn means that the actual emitter density is lower than the nanotube density by as much as several orders of magnitude \cite{4,19}. One could hence suppose that a film with nanotubes of small diameter will necessarily be more efficient as compared to a film with nanotubes of larger diameter, since the former incorporates more efficient emitters. However,
Fig. 4 reveals that the field emission is not governed solely by the diameter of the smallest tubes. If this were the case, the emission fields should decrease further for $x > 5$ mm, as the diameter of the smaller tubes decreases down to $\sim$ 30 nm. This shows that the field amplification is not given only by the geometrical shape of the emitting nanotubes. As soon as several emitters are assembled to form a film emitter, screening between the emitters will become significant even when the distance between emitters is larger than their height. In fact, electrostatic calculations reveal that the field amplification drops rapidly for intertube distances smaller than twice the tube height \cite{12,20}. Since the density of emitters increases with decreasing distance, there is an optimum distance for a maximal emitted current density that amounts to 1-2 times the tube height \cite{12,20}.

In the case of Fig. 1d, the smaller tubes (i.e., the most probable emitters) are surrounded by higher and thicker tubular structures which critically influence the field amplification \cite{12,13}. This demonstrates clearly that the field amplification of a film emitter is determined by both the geometrical shape of the emitters and their surroundings, as the presence of nearby objects screens the applied electric field and provokes a decrease in the effective field amplification. In our case, the presence of the larger tubes far from the filament is detrimental to the field emission, as the emission fields are higher in spite of the smaller minimal diameter of the tubes. This shows again the necessity of controlling precisely the morphology of the film to optimize the emission properties.

The patterned C:N nanotubes can be compared with continuous films obtained by the same method \cite{21}. The best emission obtained on a continuous film was characterized by $E_{i} = 3.8$ V/$\mu$m, $E_{thr} = 7.8$ V/$\mu$m with $\beta = 675$, which is a slightly inferior performance with respect to the structured sample at $x = 5$ mm ($E_{i} = 3.7$ V/$\mu$m, $E_{thr} = 7.2$ V/$\mu$m and $\beta = 750$). Further taking into account that only 10\% of the substrate surface are covered with nanotubes in the case of the structured film, and that the continuous films showed poor homogeneity, we conclude that the structured deposition of the catalyst allows a far better control over the emission properties of the sample.

We address finally the mechanism of emission, and in particular the role of the incorporated nitrogen. The C:N nanotubes obtained by HF-CVD compare quite well with patterned carbon nanotube films obtained by thermal CVD of acetylene in a tubular flow reactor \cite{13}. The pure carbon nanotube samples were also structured by microcontact printing using a Fe catalyst with patterns identical to the ones used here. The best nanotube sample had $E_{i} = 1.8$ V/$\mu$m, $E_{thr} = 3.3$ V/$\mu$m, with $\beta = 1208$. Carbon nanotube films with slightly inferior properties showed performances nearly identical to the best C:N nanotubes observed here. For example, one carbon nanotube film had $E_{i} = 3.6$ V/$\mu$m, $E_{thr} = 7.0$ V/$\mu$m with $\beta = 850$) which is nearly identical to the C:N nanotube film at $x = 5$ mm. It is hence very probable that the factors governing the field emission are very similar for the pure carbon and C:N nanotubes, and that the incorporated nitrogen plays little, if any, role in the field emission. This is not very surprising as the amount of incorporated nitrogen is quite small. We suspect however that the field emission properties may well be influenced by higher nitrogen concentrations. In that respect, it is interesting to compare our results with the ones
obtained by Ma et al. \cite{14}. They realized aligned nitrogen-containing carbon nanotubes with up to 10\% of nitrogen by microwave plasma-assisted CVD \cite{14}. A threshold field of $E_{thr} = 2.8$ V/$\mu$m was obtained, which is significantly lower than the values obtained in this work ($E_{thr} \geq 7.2$ V/$\mu$m). This difference is probably due in part to the higher nitrogen content, but care has to be taken in this comparison because the experimental setups and the structure of the tubes was signiffcantly different. First, Ma et al. used larger interelectrode distances (300 $\mu$m vs. 125 $\mu$m in our case) which decreases the measured threshold fields (this was also observed in this work, as can be seen in Fig. 4d around $x = 2.5$ and $7.5$ mm, where the field emission was measured for two different interelectrode distances). Second, the tube ends were opened in their case and closed in our case, which may also induce significant differences in the field emission properties \cite{1}.

As for the actual emitting structures, we argue that the emission originates from the tip of the tubes, and not from the needle-like structures decorating the body of the tubes, at least for the tubes of small diameter. The latter structures are very thin  radius of curvature $r \approx 10$ nm), but of small height $h$ ($\sim$ 50 nm) and densely packed on the body of the tube. The ability of a single structure to amplify the applied field, which can be estimated as $h/r$, is typically 10 for a single needle, whereas it easily reaches 40 for a tube as a whole. As field emission is an extremely selective process, we can conclude that the emission will originate from the tip of the tubes, at least for small diameter tubes (50 - 200 nm). This means also that the catalyst particles that are present at the tips probably influence the field emission properties, although it is difficult to say at present to what extent. For larger tube diameters  especially just underneath the filament), the probability that the needles contribute to the field emitted current becomes significantly larger.

\section*{Conclusion}

We were able to show that the deposition conditions in HF-CVD influence critically the morphology of patterned C:N nanotube films, which is in turn directly correlated to the field emission performances. The most promising range for field emission applications is found to be at $x = 4 - 6$ mm, corresponding to $T_{sub} =  730 - 750^{\circ}$C. Once the optimal conditions have been determined, a homogeneous and controlled film growth can be obtained by adjusting $r$ and $T_{fil}$ in a multiple parallel filament arrangement. This ability to tailor the tube diameter by HF-CVD could prove to be a decisive advantage to produce emitting structures
of high quality.

\section*{Acknowledgements}

We are grateful to the Centre Interdepartemental de Microscopie Electronique of EPFL (CIME-EPFL) for access to SEM and TEM facilities. The work was partly supported by the Swiss National Science Foundation.

\end{document}